\documentstyle[twocolumn,aps,epsf,prb]{revtex}
\begin{document}
\title{Anomalous Properties of Weak-Links-Containing
Superconductors: Flicker Noise}
\author{Sergei A. Sergeenkov}
\address{Bogoliubov Laboratory of Theoretical Physics,
Joint Institute for Nuclear Research, Dubna, Moscow Region, Russia}
\address{
\centering{
\medskip\em
\begin{minipage}{14cm}
{}~~~
Low-frequency magnetic noise spectral density is calculated within the
superconductive glass model. The model predicts the existence  of
both white noise and flicker-like noise $f^{-\alpha }$  with $\alpha $
strongly dependent on  applied  magnetic  field  and
temperature. It is  shown,  in  particular,  that $\alpha $
increases with field from $0.5$ to $1$, and that in  the
critical region (near $T_{c}(H)$) the $1/\sqrt{f}$ law  should
prevail in the noise spectrum.
{}~\\
\medskip
{}~
\end{minipage}
}}
\maketitle
\narrowtext

\section{Introduction}

Probably the most intriguing, perplexing, and  still
unresolved problem in SQUIDology is the  problem  of
noises (see, e.g.,~\cite{ref1} and  references  therein).  As
was indicated by  Donaldson~\cite{ref2}, all high-$T_{c}$ $RF$
SQUIDs suffer from excessive $1/f$ noise contributions
at low frequencies generated by a dense  network  of
superconducting   loops ($\simeq 10 \mu m$ in diameter)
interrupted by Josephson junctions. According to
Gross et al.~\cite{ref3}, the temperature and field
dependencies of the flux noise power spectrum
in $DC$ SQUIDs strongly correlate with the similar
behavior of the critical current (magnetization) in
weak-link-containing systems. Gabovich et al.~\cite{ref4}
observed the flicker noise in ceramic superconductor
$BaPb_{0.75}Bi_{0.25}O_{3}$ in the range $20Hz\le f\le 500Hz$ with the
noise spectral density $S_{m}(f)$ well-fitted by the law
$f^{-\alpha }$, where $\alpha $ increases with applied field.

On the other hand, as was shown in our previous
paper~\cite{ref5}, the superconductive  glass (SG) model
enables us to describe quite satisfactory
experimentally observed anomalies of critical
currents and current-voltage characteristics of
weak-link-containing superconductors (such as
ceramics and defected single crystals).

In this paper we consider via the $SG$ model the
low-frequency noise properties of granular
superconductors due to the long-time magnetization
fluctuations of the sample.

\section{The SG Model}

As is well known~\cite{ref5,ref6} the  Hamiltonian of the $SG$
model in the pseudospin representation has a form
\begin{equation}
{\cal H}= -Re \sum  J_{ij}S^{*}_{i}S_{j},
\end{equation}
where
\begin{equation}
J_{ij} = J(T)e^{iA_{ij}}, \quad S_{i} = e^{i\phi _i},
\end{equation}
with
\begin{equation}
A_{ij} = \pi H (x_{i}+x_{j})(y_{i}-y_{j})/\Phi _{0}.
\end{equation}
The model (1) describes the interaction between
superconductive clusters (with phases $\phi _{i}$) via
Josephson junctions (with an energy $J(T)$) on $2D$
disordered lattice (with cluster coordinates
$(x_{i},y_{j})$) in a frustrated external magnetic field ${\bf H}
= (0,0,H)$. The field is normal to the $ab$-plane where
a glass-like picture of high-$T_{c}$ oxides is
established~\cite{ref5,ref6}.

According to the fluctuation-dissipation theorem  we
calculate the enhancement of excess magnetic noise
density within the $SG$ model:
\begin{equation}
S_{m}(f)=\int^{\infty }_{0}dt\cos (2\pi ft)<M(t)M(0)>
\end{equation}
where
\begin{eqnarray}
M(t)&=& M(H)D(t), \\ \nonumber
M(H)&=&-M_{0}(H /H_{0})(1+H^{2}/H^{2}_{0})^{-3/2},
\end{eqnarray}
and
\begin{equation}
H_{0} = \Phi _{0}/2S, \quad M_{0}$ = SNJ$/\Phi _{0}.
\end{equation}
Here $S^{1/2}$ is an average grain size.  The correlator
$D(t)=D_{\hbox{ii}}(t)$ determines all the dynamical properties
of the model and is defined as follows:
\begin{equation}
D_{ij}(t) = <S_{i}(t)S^{*}_{j}(0)>
\end{equation}
The bar denotes the Gauss-like averaging over random
grain  coordinates $(x_{i},y_{j})$ with a mean square
deviation $S/\pi $. In the  mode-coupling approximation one obtains
the self-consistent equation in correlator $D(t)$:
\begin{equation}
d^{2}D/dt^{2}+\Omega ^{2}D(t)+\int_0^t dt'K(t-t')[dD(t')/dt']=0
\end{equation}
The Laplace transform of the memory kernel $K(t)$ in
the pair spin approximation has the form~\cite{ref5}:
\begin{eqnarray}
K_{ij}(z)&=& \Omega ^{2}F(T,H_{0})[D_{ij}(z)\\ \nonumber
  &+&i\int _0^\infty dt e^{izt}\sum^{N}_{kl}D_{ik}(t)D_{kl}(t)D_{lj}(t)]
\end{eqnarray}
where
\begin{eqnarray}
F(T,H)&=&J^{2}(T,H)/T^{2},\\ \nonumber
J(T,H)&=&J(T)/(1 + H^{2}/H^{2}_{o})^{1/2}.
\end{eqnarray}
Here $\Omega  = 2e^{2}RT/h^{2}N$ is a characteristic frequency  of
the system (1), $N$ is the number of superconducting
grains (or weak links).

Let us consider the solution of system (6)-(7)
for equilibrium relaxation in the form:
\begin{equation}
D(t) = L + (1 - L)\Phi (t) .
\end{equation}
Here we have introduced the nonergodicity (order)
parameter $L(T,H)$ which is defined via correlator
$D(t)$ as~\cite{ref5,ref6}:
\begin{equation}
L(T,H) = \lim_{t\to \infty} D(t)
\end{equation}
From~\cite{ref5} it is known that the transition temperature
$T_{c}(H)$ to the $SG$ state is determined by the equation
$L(T_{c},H) = 0.$ Taking into account eq.(9) from eqs.(6)
and (7) we obtain the equation for $L$ :
\begin{equation}
L^{3} - L^{2} + L - (1 - T^{2}/T^{2}_{c}(H)) = 0.
\end{equation}
Using as Ansatz a power-like relaxation law:
\begin{equation}
\Phi (t) = (1 + t/\tau )^{-n},
\end{equation}
for the exponent $n(T,H)$ one obtains near $T_{c}(H)$:
\begin{equation}
n = 1/2 + L/(1 - L)
\end{equation}

\section{Flicker Noise in the SG Model}

To obtain the magnetic  noise  density  due  to  the
magnetization fluctuations within the $SG$ model we
use the power-law Ansatz (11). Taking into account
eqs. (4), (8), (11), and (12) from eq.(3) one gets
finally:
\begin{equation}
S_{m}(f) = S_{o}(H)[L\delta (f)+(1-L)\tau\Gamma (\alpha )(f\tau)^{-\alpha }] ,
\end{equation}
where
\begin{equation}
\alpha (T,H)=1-n(T,H), \quad S_{o}(H)=M^{2}(H).
\end{equation}
Thus, according to eq.(13), the magnetic
fluctuations due to redistribution of  the  magnetic
flux  between  Josephson  network  loops   lead   to
emergence  of  the  white   noise   with   amplitude
$S_{o}(H)L(T,H)$ and the flicker noise $f^{-\alpha }$ with amplitude
$S_{o}(H)(1-L)$. Near $T_{c}(H)$, as it follows from  eq.(10),
the order parameter $L\simeq 1-[T/T_{c}(H)]^{2}$. In  turn,  the
phase boundary $T_{c}(H)$ is determined by  the  equation
$F(T_{c},H)=1$ (see eq.(7)), so that near $T_{c}(H)$ the white
noise contribution is strongly depressed in
comparison with the flicker noise one. Moreover,  in
view of eqs.(12) and (14), the power exponent $\alpha $ will
increase with field $H$ from $\alpha =0.5$ (at low fields
$H/H_{o}\ll 1)$ to $\alpha =1$ (at higher fields $H/H_{o}\ge 1,$ when the SG
state is established throughout the sample). It  is
important to stress that $\alpha (T_{c},H)=0.5$,  i.e.  in  the
critical region $1/\omega$  noise  should  prevail  in  the
fluctuation spectrum.


\newpage
\begin{thebibliography}{99}
\bibitem{ref1} M.J. Ferrari, this volume
\bibitem{ref2} G.B. Donaldson, Cryogenics {\bf 28}, 668 (1988).
\bibitem{ref3} R. Gross, P. Chaudhari, M. Kawasaki, M. Ketchen,
and A. Gupta, Physica C {\bf 170}, 315 (1990).
\bibitem{ref4}
A.M. Gabovich, D.P. Moiseev, V.M. Postnikov, V.A. Kulikov, and L.V.
Matveets, Physica B {\bf 165\&166}, 1165 (1990).
\bibitem{ref5} S.A. Sergeenkov, Physica C {\bf 167}, 339 (1990).
\bibitem{ref6} S.A. Sergeenkov, Z.Phys.B {\bf 82}, 325 (1991).
\end{thebibliography}
\end{document}